# Lineage-specific transposons drove massive gene expression recruitments during the evolution of pregnancy in mammals

Vincent J. Lynch[1*], Mauris Nnamani[1], Kathryn J. Brayer[1], Deena Emera[1,7], Joel O. Wertheim[2], Sergei L. Kosakovsky Pond[3], Frank Grützner[4], Stefan Bauersachs[5], Alexander Graf[5], Aurélie Kapusta[6], Cédric Feschotte[6], Günter P. Wagner[1*]

1) Yale Systems Biology Institute and Department of Ecology and Evolutionary Biology, Yale University, New Haven, CT, 06511 USA
2) Department of Pathology, University of California, San Diego, 220 Dickinson St. Suite A, San Diego, CA 92103
3) Department of Medicine, University of California, San Diego, 220 Dickinson St. Suite A, San Diego, CA 92103
4) The Robinson Institute, School of Molecular and Biomedical Sciences, University of Adelaide, Adelaide, South Australia 5005, Australia
5) Laboratory for Functional Genome Analysis (LAFUGA), Gene Center, LMU Munich, Feodor Lynen Str. 25 81377 Munich, Germany
6) Department of Biology, University of Texas, Arlington, TX 76019
7) Present address: Department of Genetics, Yale University School of Medicine, New Haven, CT 06510

* Corresponding authors: Yale Systems Biology Institute and Department of Ecology and Evolutionary Biology, Yale University, New Haven, CT 06520, USA Phone: 203-737-091, email:vincent.j.lynch@yale.edu, gunter.wagner@yale.edu

A major challenge in biology is explaining how novel characters evolve, however, the molecular mechanisms that underlie the emergence of evolutionary innovations are unclear. Here we show that while gene expression in the uterus evolves at a slow and relatively constant rate, it has been punctuated by periods of rapid change associated with the recruitment of thousands of genes into uterine expression during the evolution of pregnancy in mammals. We found that numerous genes and signaling pathways essential for the establishment of pregnancy and maternal-fetal communication evolved uterine expression in mammals. Remarkably the majority of genes recruited into endometrial expression have *cis*-regulatory elements derived from lineage-specific transposons, suggesting that that bursts of transposition facilitate adaptation and speciation through genomic and regulatory reorganization.

An enduring problem in biology is explaining how evolutionary novelties originate and evolve (*1-3*). Comparative studies indicate that changes in the timing, location, or level of gene expression influence the development and evolution of morphology (*4-6*). Although it is clear that gene regulatory evolution is ultimately responsible for phenotypic differences between species, the molecular mechanisms that underlie the emergence of innovations and their magnitude are just beginning to be revealed. For example major evolutionary transitions have been associated with large-scale reorganization of gene expression (*7-9*). However, the lack of detailed studies of global transcriptome changes from species that span the origin of innovations is hampering our understanding of their evolution. Here we present an in-depth analysis of gene regulatory changes underlying a major evolutionary innovation – mammalian pregnancy.

Pregnancy is one of the defining evolutionary innovations of mammals. While Eutherian (placental) mammals have long gestations with strong maternal recognition of pregnancy, invasive placentas, and a uterine endometrium that can accommodate implantation of the blastocyst, their sister group, the marsupials, have shorter gestations, weak maternal recognition of pregnancy, and generally lack implantation and invasive placentas (*10, 11*). Monotremes, the sister-group of placental and marsupial mammals, are oviparous but the developing intrauterine embryo is nourished from maternal secretions delivered through a primitive yolk-sac placenta (*12*). Thus extant mammals span stages in the evolution of pregnancy, and are an excellent system to study how the transcriptome evolved during a major evolutionary innovation.

**Thousands of genes were recruited into endometrial expression in mammals**

To study how gene expression in the uterus evolved during the origin of pregnancy in mammals, we sequenced the endometrial transcriptomes from six pregnant Eutherian mammals (human, dog, cow, horse, pig, and armadillo), a pregnant marsupial (short-tailed opossum), and a gravid monotreme (platypus) using high-throughput Illumina sequencing (mRNA-Seq). These datasets were integrated with existing genome-wide endometrial gene expression data from Rhesus monkey (*13*), mouse (*14*), wallaby (*15*), and chicken (*14*). Our expression data matrix comprised 8,897 one-to-one (1:1) orthologs across twelve species, including all major Eutherian lineages (Euarchontoglires, Laurasiatheria, and Xenarthra), the two major lineages of marsupials (Ameridelphia and Australidelphia), the egg-laying monotreme (platypus), and a Sauropsid (chicken) as an outgroup.

While multiple processes such as gene duplication and invention, the emergence of novel non-coding RNAs, and quantitative changes in gene expression levels likely have played a role in the evolution of pregnancy, here we focus on the role of gene recruitment into and loss from endometrial expression. Using a model for calling gene expression presence/absence in mRNA-Seq data that has been validated with ChIP-Seq for chromatin marks associated with transcriptionally active and repressed chromatin (*16, 17*) and parsimony to map gene expression recruitment and loss events, we found that 1891, 1285, and 878 genes were recruited into endometrial expression in the stem-lineage of Mammalia, Theria, and Eutheria, respectively, while 339, 134, and 344 genes were lost from endometrial expression in these lineages (Fig. 1A). Gene recruitment and loss events occur at a relatively clock-like rate (6.67 recruitment-loss events per million years;

$R^2$=0.836; Fig. 1A), however, the Mammalian, Therian, and Eutherian stem-lineages, as well as human and Rhesus monkey, have dramatically higher rates of gain and loss than other lineages. These results suggest that while the rate of evolutionary gene recruitment is relatively constant, it is punctuated by periods of rapid change during major evolutionary transitions in reproductive biology, such as the evolution of matrotrophy in Mammalia (*12*), viviparity in Theria (*10*), invasive placentation/maternal recognition of pregnancy in Eutheria (*10*), and interstitial implantation in primates (*18*).

We found that the most abundant KEGG pathway annotations for recruited genes were metabolism, signaling, extra-cellular matrix/cell adhesion, and transcriptional regulation (Fig. 1B). To determine if recruited genes play essential roles in pregnancy, we annotated genes based on their knockout phenotype in mice and their association with disease related to pregnancy in humans. We found that genes recruited in the stem-lineage of Eutherians were enriched 5.5-fold for female infertility in knockout mice, and 5-fold for pregnancy-related disease genes in humans compared to ancestrally expressed genes (Yates corrected $\chi^2$, $P<2.5\times10^{-6}$; Fig. 1C). However, this enrichment was specific to Eutherian recruits and was not found for genes recruited earlier in mammalian evolution, suggesting the Eutherian-recruited genes are preferentially associated with implantation, decidualization, or other mechanisms important for establishing pregnancy. Interestingly, while previous studies found that genes expressed in the placenta evolved rapidly during the evolution of pregnancy in early mammals (*19, 20*), we found that the prevalence of positive selection in recruited genes was not significantly different than ancestrally expressed genes (Fig. 1D).

**Recruitment of cAMP signaling in placental mammals**

Progesterone is a key regulator of decidualization, implantation, and the establishment and maintenance of pregnancy. Although progesterone acts though multiple pathways to exert its effects, progesterone-dependent decidualization of the endometrium is mediated by the second messenger 3'-5'-cyclic adenosine monophosphate (cAMP) and the protein kinase A (PKA) signaling pathways (Fig. 2A) (*21, 22*). To trace the origin of cAMP/PKA signaling in endometrium we reconstructed the expression history of proximate effectors of cAMP/PKA signaling including G-protein coupled receptors (GPCRs), adenylate cyclases (ADCYs), cAMP phosphodiesterases (PDEs), and kinases. We found that several GPCR and ADCY and numerous kinase genes were recruited into endometrial expression in mammals suggesting that the recruitment of signaling pathways played an important role in the evolution of pregnancy perhaps to establish maternal-fetal communication.

Next, we examined the expression level of enzymes that synthesize (adenylate cyclases) and degrade (phosphodiesterases) cAMP. We found that the expression of adenylate cyclases (*ADCY1-10*) was significantly higher in placental mammals than either opossum or platypus while the expression of cAMP-specific phosphodiesterases (*PDEs*) was similar across all species (Fig. 2B). These results suggest that the cAMP-signaling pathway was recruited into endometrial cells by increasing the rate of cAMP synthesis by adenylate cyclases rather than decreasing the rate of degradation by phosphodiesterases. Indeed, previous studies have shown that decreasing the rate of cAMP synthesis severely impairs decidualization, while decreasing the rate of cAMP degradation potentiates decidualization (*23*).

**The anti-estrogenic action of progesterone evolved in placental mammals**

An essential role of cAMP/progesterone signaling in the establishment of pregnancy is the down-regulation of estrogen signaling. In mice, the transcription factor HAND2 plays a central role in this process by mediating the anti-estrogenic action of progesterone at the time of implantation (*24, 25*). In the proliferative phase of the reproductive cycle, estrogen acts through ESR1 in stromal cells to increase the production of fibroblast growth factors (FGFs), which in turn serves as paracrine signals leading to sustained proliferation of epithelial cells. Active estrogen signaling maintains epithelial expression of mucin 1 (MUC1), a cell surface glycoprotein that acts as a barrier to implantation. During the receptive phase of the cycle, however, progesterone activates HAND2 expression in the endometrial stroma, inhibiting production of fibroblast growth factors (FGFs), suppressing epithelial proliferation and antagonizing estrogen-mediated expression of MUC1, thereby promoting uterine receptivity to implantation (Fig. 3A)(*24, 25*).

To identify when the anti-estrogenic HAND2 signaling pathway was recruited into the endometrium, we characterized the expression level of its key components in our transcriptomic data. We found that *HAND2* expression was recruited into the endometrium in the stem-lineage of Eutherians, coincident with the loss of *ESR1* expression during pregnancy (Fig. 3B). These results suggest that the absence of *HAND2* expression in the endometrium of non-placental mammals is associated with persistent estrogen signaling during pregnancy. Indeed, we observed that *MUC1* was expressed in chicken, platypus and opossum, as well as cow, pig and horse, but was absent or expressed at very low levels in armadillo, dog, mouse and human (Fig. 3B). Parsimony reconstruction indicates that endometrial expression of *MUC1* was lost in the stem-lineage of placental mammals, and re-established in ungulates (hoofed mammals; Fig. 3B). This pattern recapitulates the evolution of placental invasiveness, which evolved in the stem lineage of placental mammals and was secondarily lost in ungulates (*26, 27*). Hence, sustained *MUC1* expression in the endometrium seems to be a key mechanism preventing invasive placentation. Furthermore, conserved endometrial expression of *HAND2* as well as the suppression of *ESR1* in Eutherian mammals suggests that the non-invasive placenta of ungulates is secondarily derived and that sustained *MUC1* expression during pregnancy in ungulates is probably mediated through a pathway not involving *ESR1*.

Our data suggest that the role of estrogen signaling in pregnancy is an ancestral character that has been suppressed in Eutherians. To determine if estrogen signaling is active during pregnancy in non-placental mammals, we examined the phosphorylation status of MAPK1/3 and ESR1 using immunohistochemisty on endometrial sections from pregnant opossum (d.p.c. 12.5). We found strong expression of phosphorylated MAPK1/3 and ESR1 indicating that estrogen signaling pathway is active during pregnancy in the endometrium of opossum (Fig. 3C). Again, consistent with persistent estrogen signaling during pregnancy, we observed that MUC1 was strongly expressed in luminal and glandular epithelia (Fig. 3C). Together these results indicate that the recruitment of *HAND2* into endometrial expression in Eutherians played a critical role in suppressing estrogen signaling during pregnancy.

To identify *cis*-regulatory elements that may be responsible for the recruitment of *HAND2*, we used phylogenetic footprinting to search the genomic region around *HAND2* for putative regulatory elements conserved within placental mammals but that are either not conserved or not found in other species. We identified several experimentally validated binding sites for the vertebrate CTCF insulator protein flanking *HAND2*, including an upstream site that has been shown to bind CTCF in human, Rhesus macaque, mouse, and dog, but not opossum (*28*), suggesting this CTCF binding site is a Eutherian-specific insulator (Fig. 3D). Indeed this CTCF site marks a clear boundary between active and silent chromatin in this genomic region across numerous ENCODE cell lines. Consistent with the notion that this CTCF site demarcates active and repressed chromatin domains in hESC, we found that SCRG1, the most proximal gene upstream of the Eutherian-specific CTCF site, is not transcribed in the endometrium of any species we examined while genes downstream of this site including HAND2 and NBLA00301 were only transcribed in Eutherians. Thus we conclude that the evolution of a new barrier-type insulator upstream of HAND2 in Eutherian mammals protected it from epigenetic silencing, leading to its recruitment into endometrial expression.

In addition to HAND2, the aryl hydrocarbon receptor (AHR) plays an important role in suppressing estrogen-mediated proliferation by directly repressing *ESR1* transcription in response to progesterone (*29-31*), suggesting that recruitment of *AHR* into the endometrial expression may have also contributed to the evolution of uterine receptivity (Fig. 4A). Although we observed that *AHR*, its obligate cofactor *ARNT1*, and numerous *AHR* target genes were recruited into endometrial expression in the stem-lineage of Therians, we found that *ESR1* is repressed during pregnancy only in Eutherians (Fig. 4B/C), indicating that endometrial recruitment of *AHR* expression alone was not sufficient to repress *ESR1*. Thus the negative regulatory interaction between *AHR* and *ESR1* must have evolved later, in the stem-lineage of Eutherians.

To identify potential AHR responsive *cis*-regulatory elements of *ESR1* that may have evolved in the stem-lineage of Eutherians, we searched the genomic region around *ESR1* for predicted high affinity AHR binding sites conserved within Eutherian mammals but absent in other species. We identified a several putative AHR/ARNT binding sites including one that was contributed by a Eutherian-specific LINE L2a transposable element inserted in the third intron of *ESR1,* suggesting that this region may contain an AHR responsive repressor element acquired by the Eutherian ancestor (Fig 4D). To test if AHR represses *ESR1*, we treated estrogen-primed hESC with the potent AHR agonist 2,3,7,8-tetrachlorodibenzo-*p*-dioxin (TCDD) and monitored *ESR1* transcript levels by quantitative real time-PCR (qRT-PCR). We found that TCDD treatment down-regulated *ESR1* expression in estrogen primed hESC approximately 50%, but repression was blocked by siRNA-mediated knockdown of *AHR* (Fig. 4E). Thus AHR is required for TCDD-mediated *ESR1* repression in endometrial cells.

Next we cloned the L2a element from the human *ESR1* intron into a luciferase reporter (pGL4.46) and assayed its ability to regulate luciferase expression in response to AHR activation. We found that wild-type constructs repressed luciferase expression ~50% in TCDD-treated hESC; however, mutating the AHR site abolished repression (Fig. 4F) suggesting direct binding of AHR to this site is required to mediate repression. To confirm binding we assayed whether AHR occupies the putative site in the L2a element in estrogen-primed hESC using ChIP-qPCR and found that TCDD treatment led to AHR and HDAC1 binding (Fig. 4G), as predicted for a functional AHR responsive repressor element. Thus *ESR1* evolved to be repressed by AHR in endometrial stromal cells through the co-option of a binding site embedded within a Eutherian-specific transposable element. Notably we observed that the same AHR/ARNT binding site is predicted in the L2a consensus (ancestral) sequence and that 23% (95/412) of AHR binding sites identified in a recent ChIP-chip study of T-47D human breast carcinoma cells (*32*) are embedded within L2a copies, suggesting that the genomic spread of this transposable element family in the Eutherian ancestor profoundly rewired the AHR regulatory network of placental mammals. These results point to a key role for lineage-specific expansion of transposable elements in the emergence of regulatory innovations underlying the evolution of pregnancy. We explore this idea further in the next section.

**The majority of endometrial regulatory elements are derived from transposons**

To gain a better understanding of how the endometrial regulatory landscape has been altered by transposition, we determined the contribution of transposable elements to 3,250 H3K27me3 marked 'poised' enhancers recently identified from decidualized hESC using ChIP-chip (*33*). We found that 2,205 (67.8%) of these peaks overlap annotated transposable elements, indicating that the majority of poised enhancers derive from transposons (Fig. 5A). While most classes of mammalian transposons are represented among the co-opted elements (Fig. 5B), we observed that SINE (MIR), LINE (L1, L2, and L3), and DNA hAT (MER5A1, MER5A, MER20, and MER5B) elements in particular were over-represented compared to their density in the human genome (SI Fig. 1).

To determine if transposable elements played a role in recruiting genes into endometrial expression, we assembled and annotated the transcriptional start and end sites of transcripts expressed in decidualized hESC using the Cufflinks program (*34*) and identified poised enhancers located within 1.5kb of the start and end sites of these genes (Fig. 5C). We found that 55.3% (470/850), 55.8% (450/806), and 51.2% (215/420) of recruited genes recruited in the stem-lineage of Mammals, Therians, and Eutherians, respectively, could be associated with a transposon-derived enhancer. However, the phylogenetic age distribution of these transposons was similar across genes recruited in (Fig. 5D). This means, for instance, that a gene recruited in the Therian lineage may be transcribed from a transposable element insertion younger than the recruitment event. These results indicate that transposons not only contribute to the origin of new enhancers but also to the turnover of existing enhancers. For example *BCL10* was recruited into endometrial expression in the stem-lineage of Therians, but the most proximal upstream enhancer is a composite of an older Eutherian-specific (L2) transposon and a much younger primate-specific LTR transposon (Fig. 6E).

Similarly we found that two previously characterized enhancers of *ITGB3*, which was recruited in the stem-lineage of Eutherians, derive from a Mammalian-specific (L2c) and a primate-specific (MER4B) transposon (Fig. 5F). Previous studies demonstrated that the regulatory functions of these enhancers are dependent on a paired NFAT/PU.1 binding site (located in the L2c) and a HoxA-10 binding site (located in the MER4B) (*35, 36*). We found that the NFAT and HoxA-10 biding sites are present in the L2c and MER4B consensus sequences, respectively, whereas the PU.1 binding site is absent from the L2c consensus. These results suggest that the L2c integrated upstream of *ITGB3* in the Mammalian stem-lineage carrying an NFAT binding-site that likely was unable to recruit PU.1 and therefore unable to function as an enhancer, subsequent nucleotide substitutions generated a functional PU.1 binding-site capable of recruiting the NFAT/PU.1 complex and acting as enhancer for *ITGB3*. Finally this ancestral enhancer was augmented by the insertion of MER4B in primates increasing the complexity of *ITGB3* regulatory control through the addition of an upstream HoxA-10 binding site.

In addition to donating enhancer and repressor elements to host genomes, transposable elements can also donate novel promoters and untranslated exons to nearby genes (*37-39*). To determine if the origin of new promoters played a role in gene recruitment we generated a genome-wide map of trimethylated histone H3 lysine 4 (H3K4me3), an epigenetic mark of active/poised promoters (*40, 41*), from decidualized hESC using ChIP-Seq. We identified 22,440 H3K4me3 peaks, 11,763 (52.4%) of which overlap transposable element annotations indicating that the majority of promoters in hESC derive from transposons (Fig. 6A). While most classes of mammalian transposons are represented among the co-opted elements (Fig. 6B), we observe that MIR, Alu, and L2 retrotransposons are over-represented compared to their density in the human genome (SI Fig. 1).

Next we intersected the transcriptional start sites of genes recruited into endometrial expression in the Mammalian, Therian, and Eutherian stem-lineages with our H3K4me3 peaks to identify the proximal promoters of recruited genes (Fig. 6C). We found that the promoters for 56%, 80%, and 51% of the genes recruited in the stem-lineages of Mammalia, Theria, and Eutheria, respectively, derive from transposable elements (Fig. 6D). This analysis reveal that transposon-derived promoters were predominantly co-opted from Mammalian, Eutherian, and Primate-specific transposable elements, while transposons specific to other lineages contributed less than 3% of transposon-derived promoters (Fig. 6C). Interestingly, as for transposon-derived enhancers, transposon insertion and co-option into proximal promoters appear to be decoupled as only about a third of the co-option events occurred coincidentally with the recruitment of a gene into endometrial expression, while the remaining transposon-derived promoters were either co-opted from earlier insertion events (i.e. transposons older than the gene expression recruitment event), or replaced pre-existing promoters by introducing a new promoter/5'-UTR sequence immediately upstream or downstream of an ancestrally expressed gene's promoter (Fig. 6E-G).

To determine the cell-type specificity of transposon-derived H3K27me3 marked poised enhancers and H3K4me3 marked promoters in hESC we intersected these peaks with H3K27me3 and H3K4me3 peaks from nine other cell types (*42*). We found that 36-51% of transposon-derived enhancers and 63-80% of transposon-derived promoters were shared between hESC and other cell-types (SI Fig. 2), suggesting that transposons were more likely to contribute endometrial-specific enhancers than promoters. Our observation that some transposons integrate with pre-existing transcription factor binding sites, for example, AHR/ANT binding sites in L2a and HoxA-10 binding sites in MER4B, may provide a mechanistic explanation for this bias because the expression of transcription factors like AHR and HoxA-10 are cell-type restricted while components of the promoter complex are broadly expressed. Thus transposons with existing binding sites for cell-type restricted transcription factors are likely to have cell-type restricted regulatory effects, which also likely limits the pleiotropic effects of their regulatory activities.

**Discussion**

Changes in gene expression are thought to be key to the origin and evolution of morphological innovations (*4-6*). Though gene regulatory evolution may be ultimately responsible for phenotypic differences between species, large-scale analyses of gene expression evolution have only recently become possible and we still know very little about the molecular mechanisms underlying origin of innovations and major evolutionary transitions. Our analyses show that the evolution of pregnancy, a defining innovation of Therian and Eutherian mammals, was associated with the recruitment of thousands of genes into endometrial expression. While quantitative changes in gene expression levels likely contributed to the evolution of pregnancy, our results show that recruitment of genes not previously expressed in the endometrium played an important role in the origin of major evolutionary innovation.

We also found that the rate of gene recruitment and loss in the endometrium is fairly constant, evolving in a clock-like fashion in most lineages, but has been punctuated by episodes of higher recruitment rate that coincide with major evolutionary leaps in mammalian reproductive biology, including the emergence of matrotrophy/placentation (Mammalia), viviparity (Theria), maternal recognition of pregnancy (Eutheria), and interstitial implantation (Primates). Remarkably, these periods of increased gene recruitment into endometrial expression are associated with the invasion of new transposable elements into the genome, followed by co-option of the transposon at many loci to form new promoters driving endometrial expression of adjacent genes. This episodic pattern suggests that periods of relative stasis in gene expression evolution may be broken by the invasion of new transposon families into the genome that are co-opted into regulatory elements, globally rewiring gene regulatory networks and facilitating rapid evolutionary change, consistent with the long-standing hypothesis that bursts of transposition –possibly triggered by environmental stress– might facilitate adaptation and speciation through genomic and regulatory reorganization (*43-46*).

Gene expression innovation during the evolution of pregnancy recruited numerous genes and signaling pathways essential for pregnancy into endometrial expression, including core components of the cAMP-signaling pathway, and transcription factors that mediate the antiestrogenic action of progesterone, as for instance HAND2 and AHR. In addition to the inhibition of estrogen signaling, suppression of the maternal immune response to the allogenic fetus is essential for the establishment of pregnancy. Indeed, Medawar noted this immunological 'problem' raised by the evolution of viviparity in vertebrates almost sixty years ago (*47*). A key mechanism that prevents maternal rejection is the production of kynuerine by the conceptus at the maternal-fetal interface (*48*). Our finding that AHR was recruited into endometrial expression in the stem-lineage of Therian mammals, coincident with the evolution of viviparity, and the recent identification of kynuerine as an endogenous tumor-promoting ligand of AHR (*49*), suggests a previously unrecognized role for endometrial AHR in establishing maternal-fetal communication and promoting maternal immunotolerance. These results also hint at a mechanistic link between the evolution of maternal immunotolerance during pregnancy and the suppression of anti-tumor immune responses during cancer progression.

Acknowledgements: This work was funded by a grant from the John Templeton Foundation, number 12793 Genetics and the Origin of Organismal Complexity (G.P.W); results presented here do not necessarily reflect the views of the John Templeton Foundation. The funders had no role in study design, data collection and analysis, decision to publish or preparation of the manuscript. This work was also funded by an NIH Training Fellowship (AI43647) to J.O.W., to the Bioinformatics Core of UCSD CFAR (AI36214), and an NIH grant (GM093939) to S.L.K.P. The generation of RNA-Seq data from bovine, porcine and equine endometrium was supported by the German Ministry for Education and Research (BMBF, FUGATO-plus, COMPENDIUM). We would also like to thank R.W. Truman

**Figure Legends**

**Fig. 1.** Evolution of the endometrial transcriptome during the origin of pregnancy in mammals. (**A**) Parsimony reconstruction of gene expression recruitment and loss in the endometrium in amniotes. Numbers above branches indicate the number of genes lost from (-) and recruited into (+) endometrial expression in the stem-lineage of Mammalia (green), Theria (blue), and Eutheria (red). Branch lengths are drawn proportional to gene expression recruitment and loss events for all lineages. Inset, regression of recruitment and loss events per branch (steps, y-axis) on branch length time (million years, x-axis) indicates that gene expression evolution is relatively clock-like for most lineages ($R^2=0.836$ excluding the stem-lineages of Mammalia, Theria, Eutheria, human and macaque), but has been punctuated by episodes of rapid evolutionary change. Colored points correspond to colored lineages in tree. (**B**) KEGG pathway annotations of recruited genes. ECM, extracellular matrix. CAM/F, cell adhesion.(**C**) Genes recruited into endometrial expression in the stem-lineage of Eutheria are over-represented for female infertility in mouse knockouts (upper, 5.5-fold), and human female infertility (lower, 5-fold) compared to other lineages and ancestrally expressed genes. *, Yates corrected $X^2$ $P<2.8\times10^{-6}$. (**D**) The proportion genes experiencing episodes of positive selection coincident with their recruitment into endometrial expression in the stem-lineages of Eutheria, Theria, and Mammalia (bright bars) is similar to the proportion ancestrally expressed genes experiencing episodes of positive selection in these lineages (dark bars), but mirrors the magnitude of gene recruitment in these lineages.

**Fig. 2.** Recruitment of cAMP generating *adenylate cyclases* (*ADCY*) in the stemm-lineage of Eutherians. (**A**) The signal transduction pathway leading to an increase in intracellular cAMP is mediated by external signals that activate G-protein coupled receptors, leading to ADCY activation. (**B**) Expression of *ADCY* and *PDE* genes in mRNA-Seq data from mammals. Data from each species are displayed as a heat map of transcripts per million (TPM), the average TPM across genes for each species is also shown. The expression of *ADCY* genes has increased dramatically in Eutherian mammals.

**Fig. 3**. The HAND2 signaling pathway was recruited in Eutherian mammals. (**A**) The gene regulatory network mediated by estrogen and *ESR1* leading to proliferation is suppressed by progesterone through activation of HAND2. (**B**) Expression of *PGR*, *HAND2*, *ESR1*, and *MUC1* genes in mRNA-Seq data from mammals indicates that HAND2 was recruited into endometrial expression in the stem-lineage of Eutherians coincident with the loss of *ESR1* and *MUC1* expression. *MUC1* expression reevolved endometrial expression coincident with the loss of invasive placentation in ungulates. Data from each species are displayed as a heat map of transcripts per million (TPM). Branches in the phylogentic tree show extant and ancestral state reconstructions of placental invasiveness: red, hemochorial placenta (highly invasive); blue, endothelialchorial placenta (moderately invasive); black,

epithelialchorial placenta (non-invasive). (**C**) Immunohistochemistry showing phosphorylated ESR1 (pESR1), MAPK1/2 (pMAPK1/2), and MUC1 expression in d.p.c. 12.5 pregnant opossum uterus compared to control (IgG). (**D**) Evolution of a Eutherian-specific CTCF binding site upstream of HAND2. The upper track shows the location of *HAND2* and *NBLA00301* on chromosome 4, followed by H3K4me3 ChIP-Seq peaks from hESC, H3K27me3 ChIP-Seq peaks from a representative ENCODE cell line (NT2-D1), ChIP-Seq validated CTCF binding sites (*28*), and the conservation of this region in vertebrates.

**Fig. 4**. The AHR signaling pathway was recruited in the stem-lineage of Theria. (**A**) The gene regulatory network mediated by AHR promotes placental invasion by down-regulating cell adhesion and suppresses *ESR1* expression. (**B**) Expression of *AHR*, *ARNT1*, *ESR1*, and previously characterized AHR target genes in mRNA-Seq data. *AHR* was recruited into endometrial expression in the stem-lineage of Therian mammals coincident with recruitment of its target genes, however, ESR1 expression is not lost in endometrium until Eutheria. (**C**) Immunohistochemistry showing AHR, and ESR1 expression in d.p.c. 12.5 pregnant opossum uterus compared to control (IgG). (**D**) Identification of a conserved AHR binding site in a Eutheirian-specific LINE L2 element in the fourth intron of *ESR1*. The upper track shows the location of *ESR1* on chromosome r, followed by conservation of the LINE L2 in Eutherian mammals (light grey, poor conservation; dark grey, strong conservation). This element is not found in the genome non-Eutherians. Inset shows the conservation of the AHR binding site as a logo. (**E**) Activation of AHR by TCDD (+) reduces ESR1 expression ~50% compared to control (-). siRNA-mediated knockdown of *AHR* reduces *AHR* expression ~90% and prevents repression of *ESR1* in response to TCDD. (**F**) The wildtype but not mutant (GCG→AAA) human LINE L2 represses luciferase reporter

expression in response to TCDD (+) compared to control (-). (**G**) Activation of AHR by TCDD (+) dramatically increases AHR and HDAC1 binding to the L2 derived AHR binding site in ChIP-qPCR compared to control (-).

**Fig. 5**. The majority of H3K27me3 marked poised enhancers in endometrial cells derive from transposable elements. (**A**) We identified 2,205 of 3,250 (67.8%) H3K27me3 peaks in hESC that are derived from transposable elements. Stacked bar chart shows the percentage of non-transposon derived (non-TE) and transpon-derived peaks (by element class). (**B**) Word cloud showing the frequency of transposable element families under H3K27me3 peaks. Bar is scaled to 221 elements. (**C**) Cartoon showing window (yellow highlighting) used to associate enhancers with genes. (**D**) Proportion of genes recruited in the stem-lineages of Mammalia, Theria, and Euthieria associated with transposon-derived enhancers (taxomonic distribution of transposons are shown). Examples of transposon derived enhancers for *BCL10* (**E**) and *ITGB3* (**F**). Tracks show RefSeq gene annotations (blue), H3K4me3 peaks (black), H3K27me3 marked poised enhancers (brown), and transposable elements (green, Mammalian-specific; yellow, Primate-specific). Additional tracks in (**E**) show the location of ChIP-Seq peaks from ENCODE cell lines for transcription factors and co-factors that overlap transposon-derived enhancers. Highlighted regions in (**F**) show sequence logos of EMSA verified transcription factor binding sites generated from Eutherian mammals (yellow background) and the sequence of that binding site in the consensus sequence of the transposon (blue background).

**Fig. 6**. The majority of H3K4me3 marked regulatory elements in endometrial cells derive from transposable elements. (**A**) We identified 22,440 H3K4me3 ChIP-Seq peaks in hESC, 11,763 (52.4%) of which overlap transposable elements. Stacked bar chart shows the percentage of non-transposon derived (non-TE) and transpon-derived peaks (by element class). (**B**) Word cloud showing the frequency of transposable element families under H3K4me3 peaks. Bar is scaled to 910 elements. (**C**) Cartoon showing window (yellow highlighting) used to associate promoters with genes. (**D**) Proportion of genes recruited in the stem-lineages of Mammalia, Theria, and Euthieria with transposon derived promoters. The taxonomic distribution of transposons is shown for transposon-derived promoters. Examples of transposon derived promoters for *RPS8* (**E**) and *ALDH7A1* (**F**) and *NCAPD2* (**G**). Tracks show RefSeq gene annotations (blue), H3K4me3 peaks (black), and transposable elements (green, Mammalian-specific; red, Eutherian-specific; yellow, Primate-specific). Additional tracks in (**F**) and (**G**) show the location of ChIP-Seq peaks from ENCODE cell lines for transcription factors and co-factors that overlap transposon-derived enhancers. CuffLinks assembled gene annotations (black) are shown above RefSeq annotations in (**G**). Note that transcription of *NCAPD2* in hESC initiates from a promoter located within MIRb.

Supporting Online Material for

# Assembly of an evolutionary novelty by gene recruitment: evolution of mammalian pregnancy

Vincent J. Lynch[1*], Mauris Nnamani[1], Kathryn J. Brayer[1], Deena Emera[1,7], Joel O. Wertheim[2], Sergei L. Kosakovsky Pond[3], Frank Grützner[4], Stefan Bauersachs[5], Alexander Graf[5], Aurélie Kapusta[6], Cédric Feschotte[6], Günter P. Wagner[1*]

1) Yale Systems Biology Institute and Department of Ecology and Evolutionary Biology, Yale University, New Haven, CT, 06511 USA
2) Department of Pathology, University of California, San Diego, 220 Dickinson St. Suite A, San Diego, CA 92103
3) Department of Medicine, University of California, San Diego, 220 Dickinson St. Suite A, San Diego, CA 92103
4) The Robinson Institute, School of Molecular and Biomedical Sciences, University of Adelaide, Adelaide, South Australia 5005, Australia
5) Laboratory for Functional Genome Analysis (LAFUGA), Gene Center, LMU Munich, Feodor Lynen Str. 25 81377 Munich, Germany
6) Department of Biology, University of Texas, Arlington, TX 76019
7) Present address: Department of Genetics, Yale University School of Medicine, New Haven, CT 06510

* Corresponding authors: Yale Systems Biology Institute and Department of Ecology and Evolutionary Biology, Yale University, New Haven, CT 06520, USA Phone: 203-737-091, email:vincent.j.lynch@yale.edu, gunter.wagner@yale.edu

**This file includes**
Materials and Methods

Fig. S1 and S2

**Materials and Methods**

*High-throughput transcriptome sequencing*: Endometrial samples from mid-stage pregnant platypus, opossum, armadillo and dog, and day 14-18 post-implantation cow, horse, and pig were dissected to remove myometrial and placental tissue, and washed in ice-cold PBS to remove blood cells. Samples of differentiated human endometrial stromal cells were cultured and differentiated with progesterone and cAMP as described below. Total RNA was extracted using the Qiagen RNA-Easy Midi RNA-extraction kit followed by on-column DNase treatment (Qiagen). Total RNA quality was assayed with a Bioanalyzer 2100 (Agilent) and found to be of excellent quality for all samples. Aliquots from the total RNA samples were sequenced using the Illumina Genome Analyzer II platform and 76-bp reads, following the protocol suggested by Illumina for sequencing of cDNA samples. Two-four biological replicates each were sequenced for all samples except platypus. We used a model-based method to classify gene as expressed based on the number of transcripts per million (TPM) in the total RNA-seq dataset. Based on this method genes at least 2 TPM were classified as "expressed".

*Parsimony reconstruction of gene expression gain/loss and functional characterization of recruited genes*: 1:1 orthologus were identified from the human (GRCh37), macaque (MMUL_1), mouse (NCBIM37), dog (BROADD2), cow (UMD3.1), horse (EquCab2), pig (Sscrofa9), armadillo (dasNov2), opossum (monDom5), wallaby (Meug_1.0), platypus (OANA5), and chicken (WASHUC2) gene builds at Ensembl. We annotated the 8,897 1:1 orthologus protein coding genes across all species. For parsimony reconstruction of gene expression gains and losses we used Wagner parsimony implemented in the pars program from PhyML (v2.4.4) and custom python scripts. Genes reconstructed to have been recruited into or lost from endometrial expression were annotated based on their KEGG pathway annotations, mouse knockout phenotypes were assessed based on phenotype data available at the Mouse Gene Informatics (MGI) database while human phenotype data were based on data available from the Human Phenotype Ontology.

*Selection tests*: For positive-selection tests, we filtered the 8,897 1:1 orthologus described to remove genes not represented by at least two non-mammalian taxa and identified 5,292 alignments from 40 tetrapod genomes suitable for additional tests. We tested for the presence of positively selected amino acid substitutions in the stem-lineages of Mammalia, Theria, and Eutheria using a branch-site random effects likelihood (BS-REL) model (*20*). The BS-REL model treats selective regimes at each branch as independent random variables, with the proportion of sites evolving under purifying ($\omega_1<1$), relaxed ($\omega_1<\omega_2\leq1$), and unrestricted ($\omega_3$) selection along individual branches, and branch-specific ω parameters inferred by maximum likelihood. A lineage is identified as positively selected whenever its $\omega_3$ is significantly greater than 1 at $p \leq 0.05$ (likelihood ratio test, corrected for multiple testing).

*Immunohistochemistry (IHC)*: Endometrial tissue from pregnant opossum (d.p.c. 12.5) were fixed in 10% neutral buffered formalin, paraffin-embedded, sectioned at 4 μm, mounted on slides, and subjected to IHC using the Pierce Peroxidase IHC Detection Kit (Cat. 36000) following the manufacturers protocol. Briefly, uterine sections were incubated at 4°C overnight with polyclonal antibodies against HAND2 (Santa Cruz SC-9409), MUC1 (Novus Biologicals NB120-15481), p-ERα (Santa Cruz SC-12915), p-Erk1/2 (also known as MAPK1/2; Santa Cruz SC-23759-R), and AHR (Santa Cruz SC-5579x).

*Cell Culture*: Human endometrial stromal cells (hESCs) immortalized with human telomerase (ATCC, Cat. No. CRL-4003), were grown in DMEM, supplemented with 5% charcoal-stripped calf-serum (Hyclone) and 1% antibiotic/antimycotic (ABAM). To induce decidualization, cells were treated with 0.5 mM 8-Br-cAMP (Sigma) and 1 µM of the progesterone analog medroxyprogesterone acetate (MPA) for 48hrs. At 80% confluency, cells were collected for gene expression analysis, transfected for siRNA mediated knockdown or luciferase assays using TransIT-LT1 (Mirus) according to the manufacturer's protocol, or harvested for ChIP.

*siRNA-Mediated Knockdown/Luciferase Reporter Assays/Chromatin Immunoprecipitation:* Endogenous AHR expression was transiently knocked-down in hESC using AHR Silencer Select Validated siRNA (Ambion, Cat. 4390824), efficacy of knock-down was assayed 48hrs after transfection using TaqMan primer/probes specific for *AHR* and *ESR1*.

The human ESR1 associated L2a was cloned into the pGL4.26 luciferase reporter vector (Promega). pGL4.26 luciferase reporter constructs (100ng) and the pGL4.74 *Renilla* luciferase controls (20ng) were transiently transfected into undifferentiated and differenatied hESC and luciferase expression assayed using the Dual-Luciferase reporter system (Progema) 48hrs after transfection. Firefly luciferase activity was normalized by *Renilla* luciferase activity. Initially cells for luciferase assays were grown in DMEM, supplemented with 5% charcoal-stripped calf-serum (Hyclone) and 1% antibiotic/antimycotic (ABAM). 10,000 cells were seeded into opaque 96-well plates and either grown in the media described above or in DMEM, supplemented with 5% charcoal-stripped calf-serum (Hyclone) and 1% ABAM supplemented with 0.5 mM 8-Br-cAMP (cAMP; Sigma) and 1 μM medroxyprogesterone acetate (MPA; Sigma).

For chromatin immunoprecipitation (ChIP), the EZ-Zyme Chromatin Prep kit (Upstate Millipore, Billerica, MA) was used following the manufacturer's protocol. Briefly, chromatin was crosslinked with 1% formaldehyde for 10m, followed by quenching with glycine and DNA fragmentation. The equivalent of $10^6$ cells was used for each immunoprecipitation. The nuclear lysate was precleared for 1h with protein G magnetic beads and incubated over night at 4°C with protein G magnetic beads, and 2 μg of ChIP validated antibodies against AHR (Santa Cruz SC-5579x), HDAC1 (Santa Cruz SC-7872x), or with species appropriate IgG as negative control. Enrichment of the L2a target was evaluated by qPCR using 1/50 of the immunoprecipitated chromatin as template and the Power SYBR Green PCR Master Mix (Applied Biosystems, Foster City, CA).

*H3K4me3 Chromatin Immunoprecipitation-sequencing (ChIP-Seq):* hESC were grown in steroid-depleted DMEM, supplemented with 5% charcoal-stripped calf-serum and 1% antibiotic/antimycotic (ABAM). At 100% confluency, cells were induced to decidualized ESC by treatment with 0.5 mM 8-Br-cAMP (Sigma) and 1μM medroxyprogesterone acetate for (Sigma) 48-72 hours. Cells were harvested and cross-linking, chromatin isolation, sonication and immunoprecipitation using an anti-H3K4me3 (Invitrogen cat# 49-1005) antibody. Briefly, between 15-150 μg of isolated chromosomal DNA was incubated with 10 μg of antibody (anti-H3K4me3) coupled to the proteinG Dynabeads (invitrogen). Antibody-bead complex was prepared following manufactures instructions. Chromatin-antibody-bead complex was incubated overnight at 4°C in 1X ChIP Dilution buffer (0.02% SDS, 2.2%

Triton X-100, 2.4mM EDTA, 33.4mM Tris pH 8.1, 334mM NaCl) supplemented with protease and phosphatase inhibitors. After incubation the complex was washed 3x with IP wash buffer (NaCl) (100mM Tris pH8.0, 500mM NaCl, 1% NP-40, 1% deoxycholic acid) followed by 2X with IP wash buffer (LiCl) (100mM Tris pH8.0, 500mM LiCl, 1% NP-40, 1% deoxycholic acid) with 3 min rotation, and once with 1ml TE buffer. Chromatin was eluted in 50mM Tris pH8.0, 10 mM EDTA, 1% SDS by incubation at 65°C with agitation. The eluted DNA was incubated at 65°C overnight to reverse the cross-links. Following incubation, the immunoprecipitated DNA was treated sequentially with RNase A and Proteinase K and was then desalted using the QIAquick PCR purification kit (Qiagen). ChIP library preparation and high-throughput sequencing were performed on an Illumina Genome Analyzer II platform by following the protocol suggested by Illumina for sequencing chromosomal DNA. Sequencing was performed by the Yale Center for Genome Analysis.

*Identification of Transposable Element Derived Promoter and Enhancers*: To Identify regulatory elements derived from transposable elements (TEs), we intersected H3K4me3 ChIP-Seq peaks from the analysis described above and previously characterized H3K27me3 marked 'poised' enhancers identified from ChIP-chip in decidualized hESC (*33*) with the RepMask 3.2.7 track at the U.C.S.C genome browser. Prior to intersection, we removed non-transposable element annotations from the RepMask 3.2.7 track. TEs were classified as falling under H3K4me3 and H3K27me3 peaks if greater than 80% of the TE was found under the respective peak.

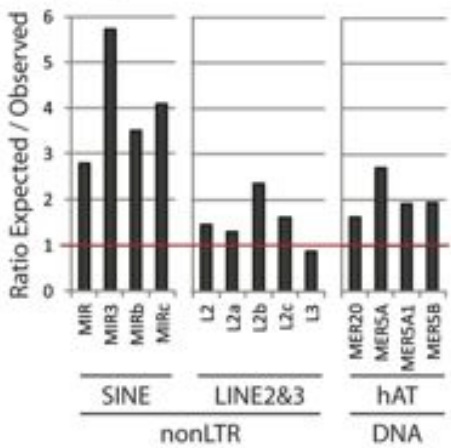
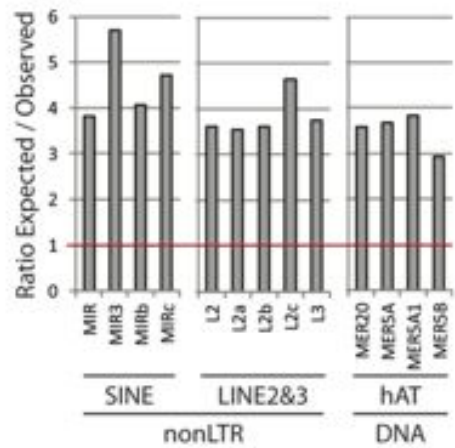
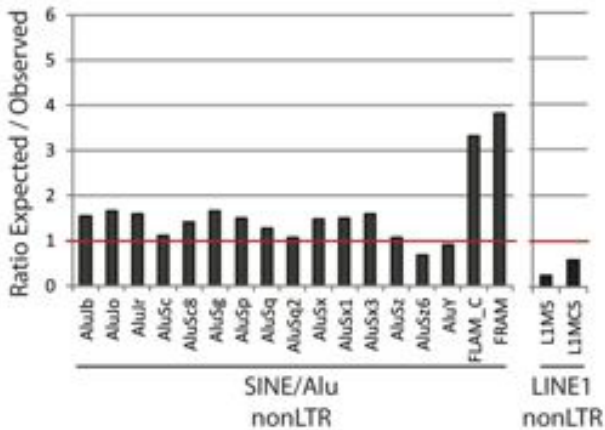
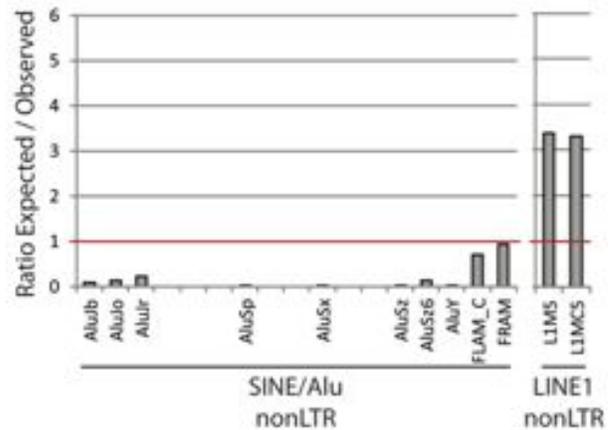

**SI Fig. 1. Over-represented TE families intersecting with peaks of H3K4me3 and H3K27me in human endometrial cells.** The expected and observed amount of DNA corresponding to each TE are calculated using Repeat Masker output (hg18, version 3.2.7). Observed values are obtained by considering overlapping TEs with H3K4me3 or H3K27me peaks. Expected values are calculated based on the overall density of each TE family in the genome according to the Repeat Masker output assuming a random distribution of TE family members throughout the genome. A ratio above 1 means that the TE is over represented in the considered set (H3K4me3 peaks in dark grey and H3K27me peaks in light grey). Only the most abundant TEs are represented (>50 hits for H3K4me3, and >20 hits for H3K27me), except for LINE1 in H3K4me3 and Alu in H3K27me (added for comparison when present in peaks).

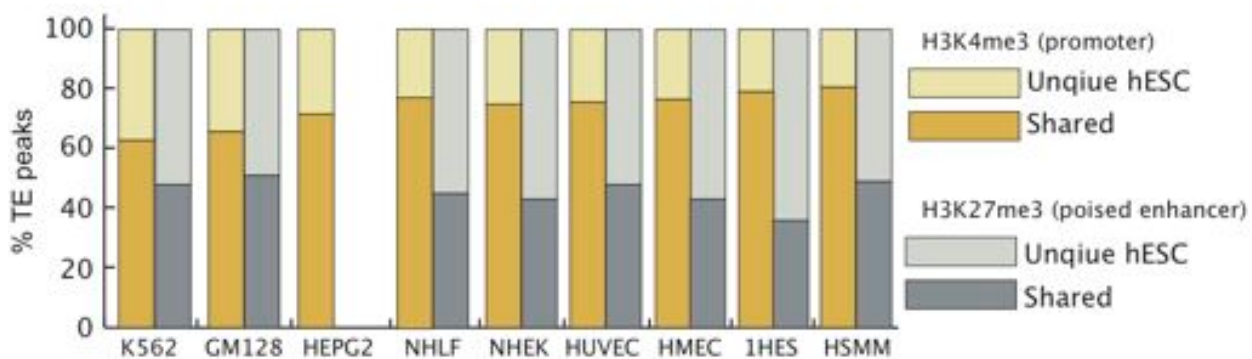

**SI Fig. 2. Cell-type specificity of transposon-derived H3K27me3 marked poised enhancers and H3K4me3 marked promoters**. H3K27me3 and H3K4me3 peaks in hESC were intersected with the same peak identified from nine other cell types (*39*). 36-51% of transposon-derived enhancers (H3K27me3) and 63-80% of transposon-derived promoters (H3K4me3) are shared between hESC and other cell-types.

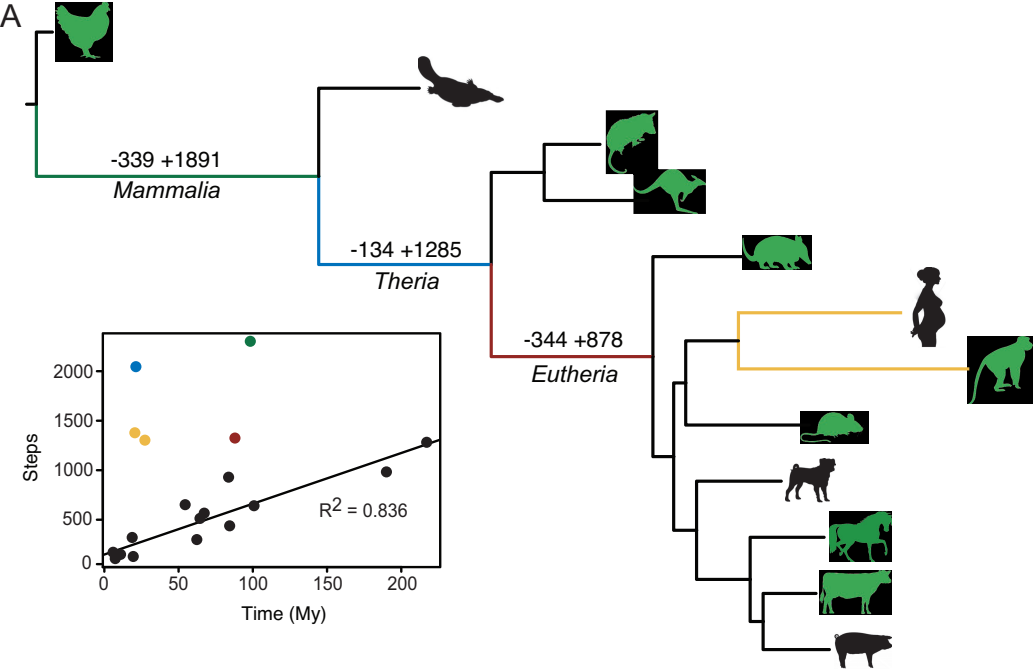
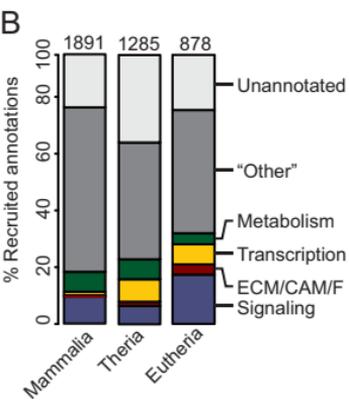
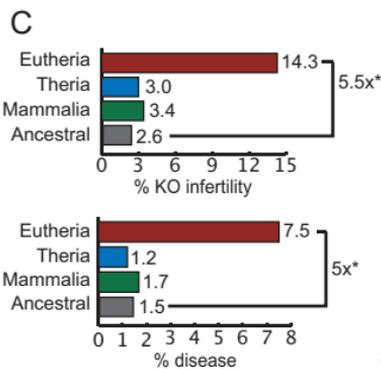
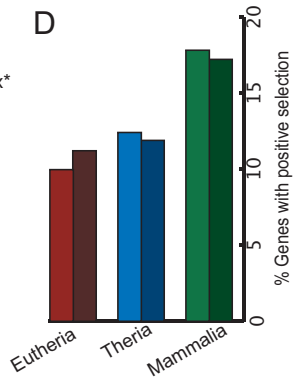

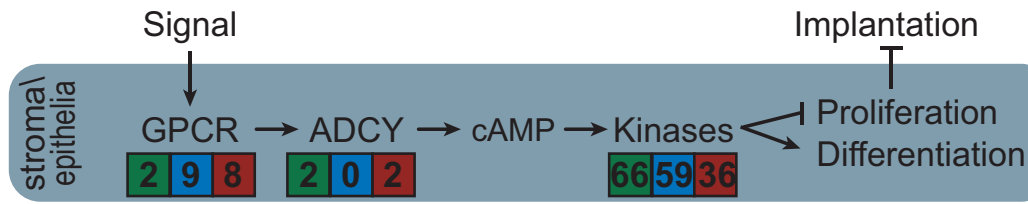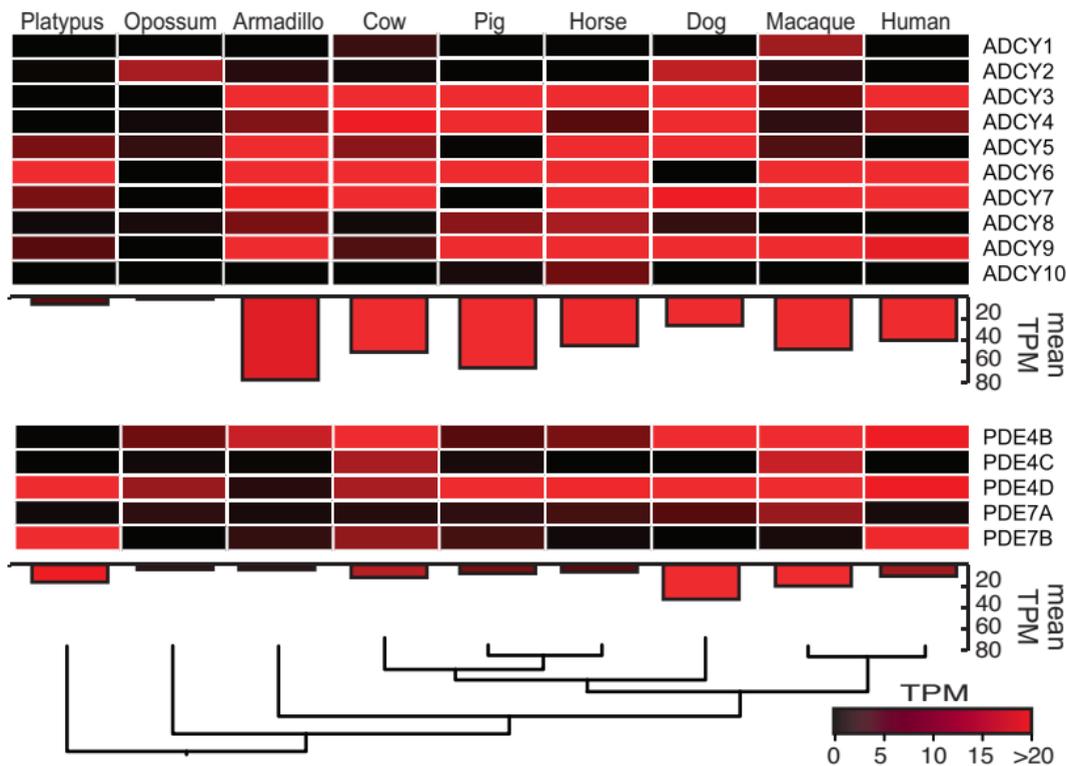

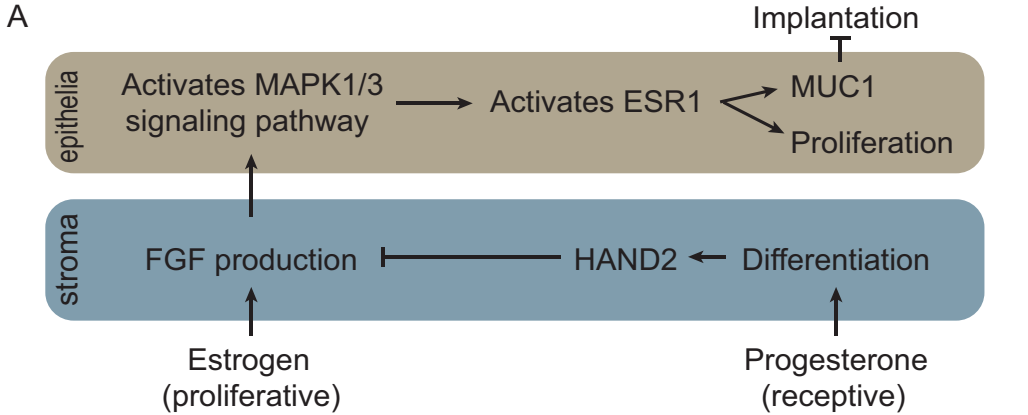

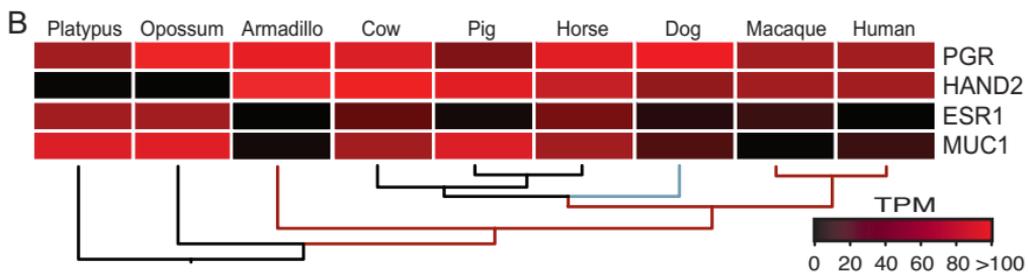

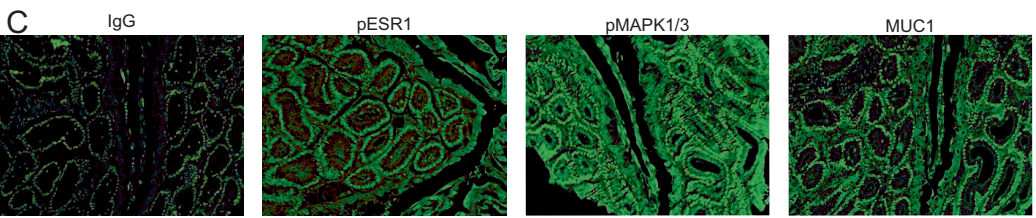

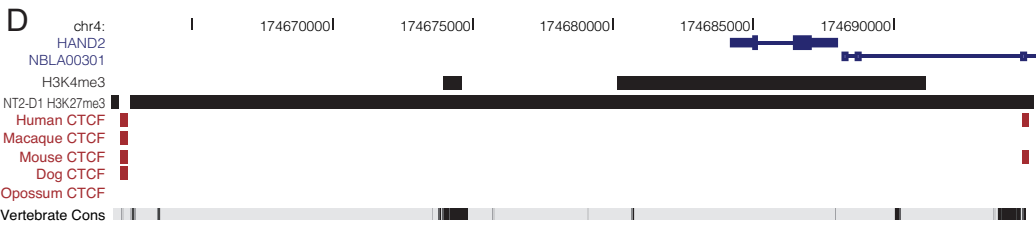

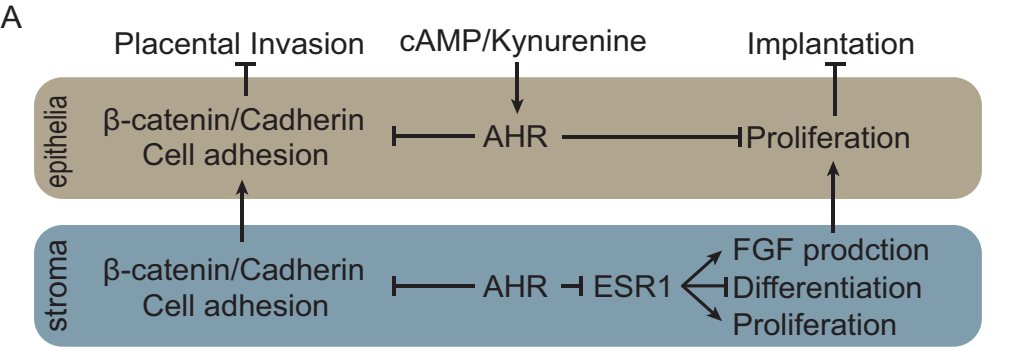
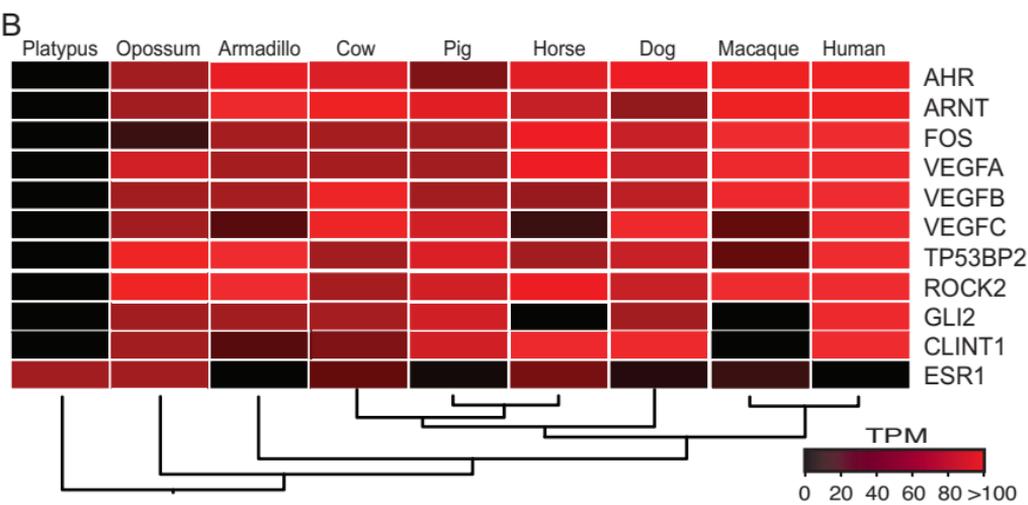
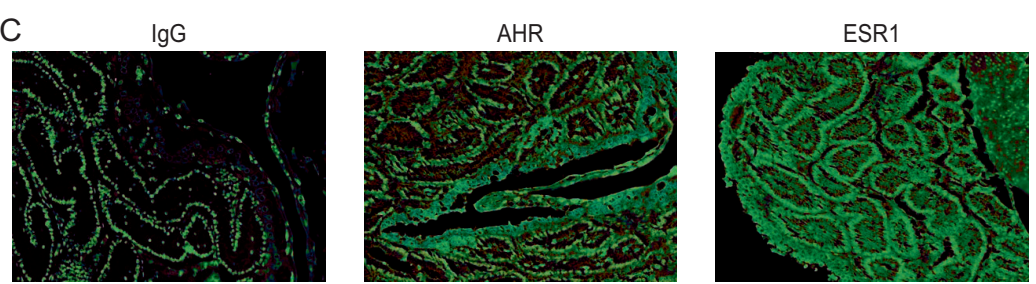
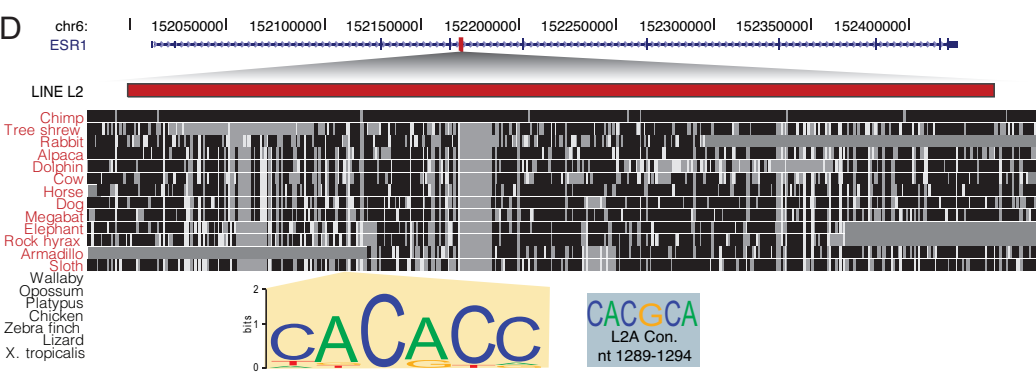
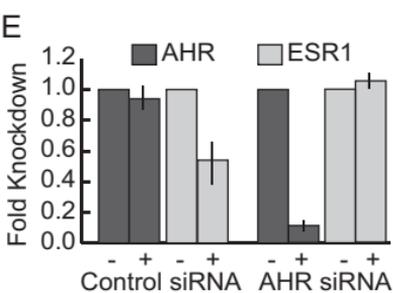
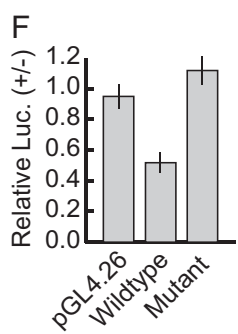
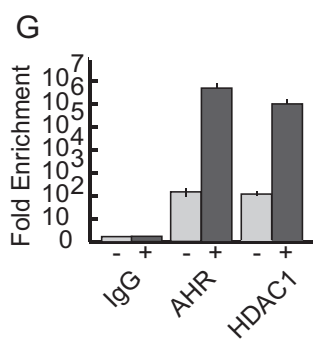

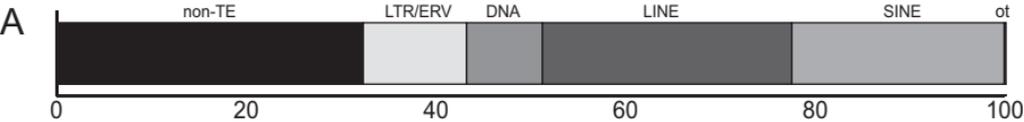
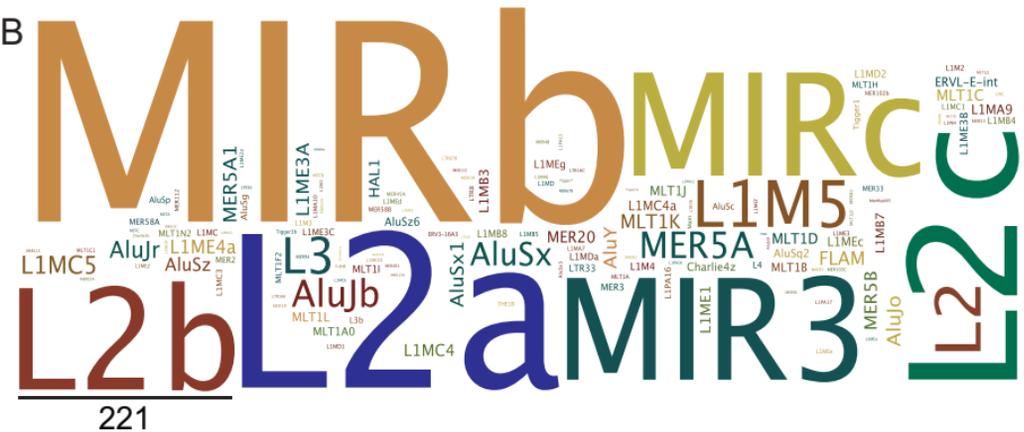
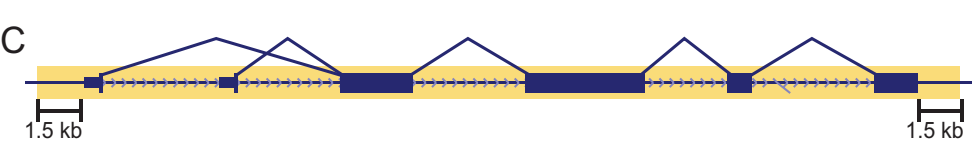
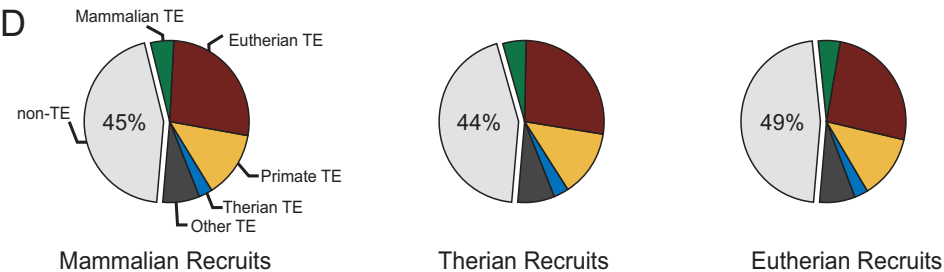
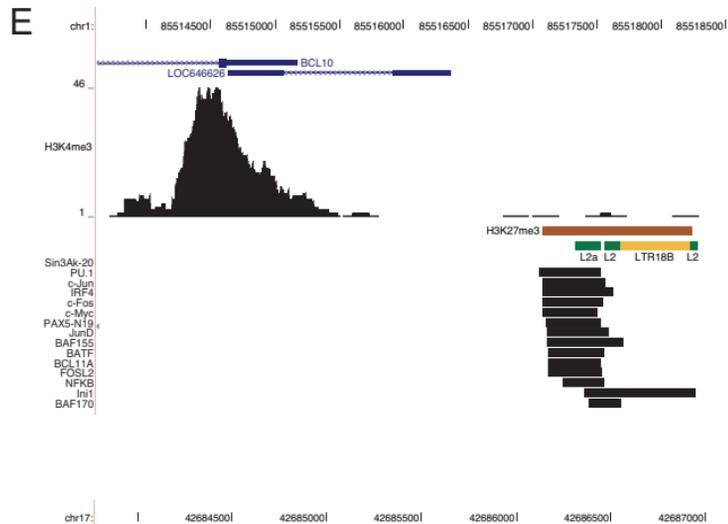
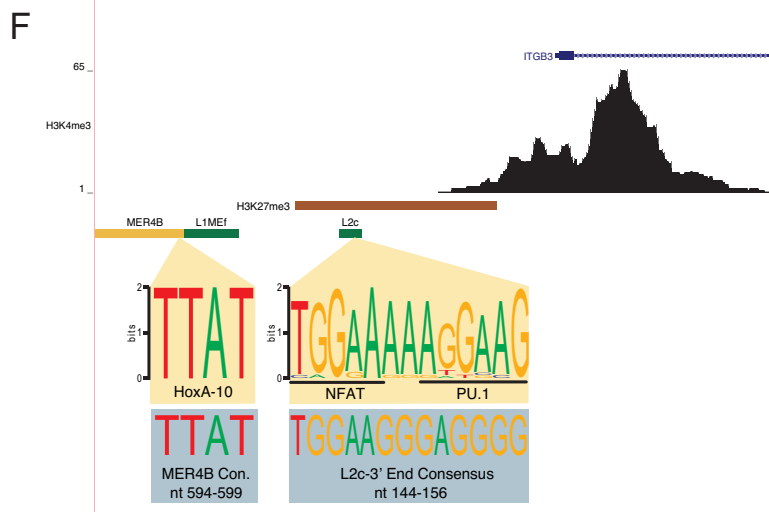